\documentstyle[referee]{mn} % for a referee version

\title{ Early GeV afterglows from gamma-ray bursts in pulsar wind bubbles}
\author[X. Y. Wang,  Z. G. Dai  and T. Lu]
{X. Y. Wang,  Z. G. Dai  and T. Lu \\
Department of Astronomy,
Nanjing University, Nanjing 210093, P. R. China \\ E-mail:
xywang@nju.edu.cn; daizigao@public1.ptt.js.cn; tlu@nju.edu.cn }

\date{Accepted ........
      Received .......;
      in original form .......}
\pubyear{2002}

\begin{document}

\maketitle

\begin{abstract}

{Gamma-ray bursts may occur within pulsar wind bubbles (PWBs)
under a number of scenarios, such as the supranova-like models in
which the progenitor pulsar drives a powerful wind shocking
against the ambient medium before it comes to death and produces a
fireball. We here study the early afterglow emission from GRBs
expanding into such a PWB environment. Different from the usual
cold GRB external medium, the PWBs consist of a hot
electron-positron ($e^+{e^-}$) medium with typical 'thermal'
Lorentz factor of the order of $\gamma_w$, the Lorentz factor of
the pulsar particle wind. After GRB blast waves shock these hot
$e^+{e^-}$ pairs, they
 will emit synchrotron radiation peaking at GeV bands. It is shown that GeV photons suffer
negligible absorption by the soft photons radiation field in PWBs.
Thus, strong GeV emissions in the early afterglow phases are
expected, providing a plausible explanation for the long-duration
GeV emission from GRB940217 detected by EGRET. Future $GLAST$ may
have the potential to test this GRB-{ PWB }interaction model.}

\end{abstract}

%
%________________________________________________________________

\section{Introduction}
Although the fireball shock scenario of GRBs and their afterglows
has been well demonstrated by multiwavelength observations of
afterglows (e.g. Wijers, Rees \& M\'{e}sz\'{a}ros 1997; Piran
1999; van Paradijs, Kouveliotou \& Wijers 2000) , the nature of
the central engine is not yet well known. There is observational
evidence for the association of long duration bursts with
star-forming regions (Bloom et al. 2002), or possibly with
supernovae (SNe; Galama et al. 1998). One class of candidates of
the central engine involves that the supernova leaves behind a
rotationally-supported supermassive or massive neutron star, which
then shrinks by shedding angular momentum via a pulsar-type wind
and finally collapses to a black hole surrounded by a  disk
(Supranova model; Vietri \& stellar 1998) or transits to a strange
star {\footnote{Strange quark matter is conjectured to be more
stable than hadronic matter (Witten 1984). The existence of
strange matter is allowable within uncertainties inherent in a
strong-interaction calculation (Farhi \& Jaffe 1984), Strange
stars, composed of this kind of quark matter, may exist and could
be born from a massive neutron star as it spins down (Wang et al.
2000)}} (Wang et al. 2000), producing the GRB fireball. The
pulsar-wind bubble (PWB) forms when the relativistic wind from the
pulsar shocks against its ambient matter and creates a pulsar
nebula like the famous Crab nebula. Though different physical
processes are involved for the death of the neutron star in these
two scenarios, the PWB and  the SN ejecta behaviors as well as the
rotation energy of the progenitor pulsar are nearly identical;
hence the two scenarios have rather similar implications to the
evolution of GRB afterglows. This class of models {\footnote {In
the literature, Fe emission features are usually interpreted in
the supranova model (Vietri \& Stellar 1998). We note however that
Fe line production is irrelevant to whether the progenitor pulsar
collapses to a black hole or converts to a strange star.}} in
which SN explosion preceded the GRB event has recently gained
support from the detections of strong Fe emission features in the
X-ray afterglow spectra of some GRBs (e.g. Piro. et al. 1999;),
particularly that of GRB991216 (Piro et al. 2000), as well as an
absorption feature in the prompt emission of GRB990705 (Amati et
al. 2000). These detections implies that a large amount
($\ga0.1M_\odot$) of pure iron is located in the vicinity
($r\la10^{16}$cm) of the GRB source (e.g. Vietri et al. 2001).
Such a surprising large iron mass is most naturally produced in a
SN explosion and the inferred distance of the SN ejecta implies
the SN event precedes the GRB by {  months to years or shorter
(see, e.g., B\"ottcher, Fryer \& Dermer 2002).}

Effects of wind nebulae resulted from the central 'pulsar' on the
GRB afterglows have recently been   examined by Inoue et al.
(2002) and K\"{o}nigl \& Granot (2002; hereafter KG) from
different points of view. Assuming the SN ejecta may be fragmented
by the plerion radiation, Inoue et al. (2002) study the inverse
Compton effect of  this radiation field on GRB external shock
decelerated by the outlying baryonic material. Different from
Inoue et al. (2002), in the model of KG, the afterglow shock wave
propagates in a  bubble composed mainly of shocked electrons and
positrons that emanate from the central pulsar. KG argue that the
pulsar wind nebula environment, bounded by the SNR shell,
 can account for the high electron and magnetic energy fractions
($\epsilon_e$ and $\epsilon_B$, respectively) inferred in a number
of afterglow sources, as the PWBs are expected to have a
significant $e^+e^-$ component and to be highly magnetized.
Following the previous treatment of the PWB structure that was
developed for the Crab and other nebulae (e.g. Rees \& Gunn 1974;
Kennel \& Coroniti 1984, hereafter KC84), and assuming the
magnetic pressure remains in approximate equipartition with the
particle pressure throughout the shocked-wind bubble, they
developed the structure model of PWB. On the other hand, a
one-zone simplified version of the shock-wind model of PWB was
developed by Chevalier (2000) and is shown to be well consistent
with the X-ray luminosity of nebulae. Guided by these works, we
attempt to give a simplified analytic description of the PWB that
may exist in the supranove-like models (section 2), and study the
early afterglow emission from blast wave propagating in the hot
$e^+e^-$ bubble (section 3) . The characteristic synchrotron
frequencies of early afterglows are much higher than that of the
afterglows in a typical interstellar medium (ISM). In section 4,
we give a brief summary.

\section{Pulsar wind bubbles}

Before collapsing, the  supermassive or massive neutron star loses
its energy through a pulsar-type wind, whose luminosity can be
estimated from the magnetic dipole formula
%\begin{equation}
$L_w=B^2 {R_{eq}^6 }\Omega^4/{6 c^3}=3\times10^{44} {\rm erg
s^{-1}}\,{B_{12}^2} {R_{eq,13}^6 }{\Omega_4^4}$
%\end{equation}
and the spin-down time, in which half of the neutron star rotation energy $\Delta{E_{rot}}\sim10^{53}{\rm erg}$ need
to be lost, is
\begin{equation}
t_{sd}=\Delta{E_{rot}} /L_w=10 \,{\rm yr}\, \Delta{E_{rot,53}}{B_{12}^{-2}}
{R_{eq,13}^{-6}} {\Omega_4^{-4}},
\end{equation}
where $B=10^{12} B_{12}{\rm Gauss} $, $R_{eq}=13 R_{eq,13}{\rm Km}
$ and $\Omega=10^{4} \Omega_4{\rm s^{-1}} $ are, respectively, the
dipole magnetic field, the typical equatorial radius and the
angular velocity of the neutron star. Due to the unknown
variations of these parameter values, we below consider $t_{sd}$
as a free parameter and parameterize the wind luminosity as
$L_w=\Delta{E_{rot}}/t_{sd}= 3\times10^{44} {\rm erg s^{-1}}{
\Delta{E_{rot,53}}} {t_{sd,1}^{-1}}$, where $t_{sd}\equiv 10{\rm
yr} \,{t_{sd,1}}$.

We assume that the PWB is elongated along the rotation axis of the
pulsar with an outer radius $R_b$ equal to that of the SN ejecta
in this direction and pulsar wind is shocked at radius $R_s$ (Rees
\& Gunn 1974; KC84; KG). The pressure force of the expanding PWB
is expected to accelerate the SN ejecta into a thin shell
(Reynolds \& Chevalier 1984) moving with a velocity
$v_{ej}=0.05-0.1 {\rm c}$ as a significant amount of $\Delta
E_{rot}$ is deposited into the bubble. Approximately, the outer
radius of the nebula at time $t_{sd}$ equals to
%\begin{equation}
$R_b=v_b t_{sd}=10^{18} {\rm cm} \, v_{b,0.1}t_{sd,1}$ ,
%\end{equation}
where $v_b=v_{ej}=0.1 {\rm c}\, v_{b,0.1}$ is the expansion
velocity of the outer edge of the PWB. Fe line emission requires
that the pre-ejected SN ejecta should be no farther than
$r\sim10^{16}{\rm cm}$. To reconcile this short Fe line-emitting
distance with the radial distance of the PWB, KG suggest that the
PWB become elongated in the polar direction because of anisotropic
mass outflow from the GRB progenitor star. We note that the
anisotropic pulsar wind power could also give rise to the
elongated configuration. In fact, some pulsar wind nebulae have
been found to show such anisotropic structure( e.g. Gaensler et
al. 2001).

The pulsar rotation energy is assumed to go into a highly
relativistic wind of the Lorentz factor $\gamma_w$. Spectral
modelling of the Crab nebula has given a current value of
$\sim3\times10^{6}$ for $\gamma_w$. However, people do not know
the initial value of $\gamma_w$ for pulsars in their early years.
A much lower initial value $\gamma_w\le10^4$ for Crab pulsar is
inferred by Atoyan (1999) when interpreting the radio spectrum of
the Crab nebula. In light of these factors, we regard that the
value of $\gamma_w$ is quite uncertain and may span a range of two
orders.  The wind luminosity consists of electromagnetic and
particle contributions with their flux ratio denoted by the
magnetization parameter $\sigma$. KC84 note that $\sigma$ must be
smaller than 0.1 in order to have a high efficiency of the wind
power into synchrotron radiation.
%Nevertheless, $\sigma\sim1$ has been inferred for the Vela synchrotron nebula (Helfand, Gotthelf \& Halpern 2001).
%On the other hand, the pulsar wind could consist of $e^+ e^-$ pairs as well as ions. Gallant \& Achterberg (1999) even
%suggested that ultra-high-energy cosmic rays might be relativistic ions in the nebula that were accelerated by
%GRB blast wave in the context of the neutron-star binaries scenario.
We assume that the pulsar wind flux is dominated by  $e^+ e^-$
pairs, so  $L_w\simeq 4\pi n_1 {\gamma_w^2} R_s^2 m_e c^3$, where
$n_1$ is the proper pair density in the wind just before the shock
front.

As the young SNR expands at a velocity $v_b\ll c$, the wind energy
excluding the part that goes into radiative and adiabatic loss
will accumulate within a volume  confined  by SNR. The
characteristic radius $R_s$ of the shock front is determined by
the balance between the ram pressure of the wind $L_w/4\pi c
R_s^2$ and the total magnetic and particle pressure $P$ in the
bubble. Since the sound speed in the shocked fluid  $c/\sqrt{3}$
is much larger than that of the SN ejecta, no pressure  gradient
in the nebular bubble can be maintained for dynamically important
time, and the bubble is virtually isobaric (Reynolds \& Chevalier
1984). At the shock front $R_s$, the wind $e^+ e^-$ particles may
acquire a power-law energy distribution of the form $N(\gamma)
\propto \gamma^{-p}$ for $\gamma\ge\gamma_m$, where $n=\int
N(\gamma) d\gamma$ is the number density of the shocked
relativistic particles, $\gamma$ is the Lorentz factor of the
particles and $\gamma_m$ is the minimum value of $\gamma$. Usually
$p>2$ is assumed to keep the total energy of the shocked particles
{ finite } and $p=2.2$ is inferred for Crab nebula from its X-ray
spectrum (e.g. Pravdo \& Serlemitsos 1981). $p\simeq 2.2$ seems
also consistent with the simulation result of the first-order
Fermi particle acceleration (Kirk et al. 2000). From the shock
jump conditions, the internal pressure immediately behind the
shock front is $p_b=n_1 m_e c^2 \gamma_w^2$ and the energy density
is $e_b=3n_1 m_e c^2 \gamma_w^2$. As in Chevalier (2000), we
assume that the energy density in the downstream region is divided
between a fraction $\epsilon_e$ into particles and a fraction
$\epsilon_B=1-\epsilon_e$ into the magnetic field, similar to the
definition in the blast-wave model of GRB afterglows (Sari et al.
1998). The plausibility  for incorporating the equipartition
assumption have been accounted for in Chevalier (2000) and KG.

With these assumptions, the magnetic field in the bubble is
\begin{equation}
B^b=(\frac{6\epsilon_B L_w}{R_s^2 c})^{1/2}={ 0.17 \,{\rm
Gauss}\,\, \alpha^{-1} \epsilon_{B,0.5}^{1/2}
t_{sd,1}^{-3/2}\Delta{E}_{rot,53}^{1/2}v_{b,0.1}^{-1} }
\end{equation}
where $\alpha\equiv R_s/R_b$ and $\epsilon_B=0.5\epsilon_{B,0.5}$.
The minimum Lorentz factor of the shocked $e^+ e^-$ is $\gamma_m=(p-2)/(p-1)\epsilon_e \gamma_w=8\times10^{3}
\epsilon_{e,0.5} \gamma_{w,5}$, where $\epsilon_e=0.5\epsilon_{e,0.5}$ and $\gamma_w=10^{5}\gamma_{w,5}$.
The synchrotron lifetime of the shocked $e^+ e^-$ with the average Lorentz factor $\bar{\gamma_e}=\epsilon_e \gamma_w$
is estimated to be
\begin{equation}
\tau_{syn}=\frac{24\pi m_e c}{\sigma_T \bar\gamma_e
(B^b)^2}=2\times10^6 {\rm sec}\,\, \alpha^2 \epsilon_{e,0.5}^{-1}
\epsilon_{B,0.5}^{-1} \gamma_{w,5}^{-1} t_{sd,1}^3 {
\Delta{E}_{rot,53}^{-1}v_{b,0.1}^{2}},
\end{equation}
where $\sigma_T$ is the Thompson cross section. Thus, for
$t_{sd}\la 10{\rm yr}$ and $\gamma_w$ ranging from $10^4$ to
$10^6$, the synchrotron lifetime of the $e^+ e^-$ in the bubble is
much shorter than the age of the bubble and the PWB is in the
strongly cooling regime for parameters of interest. In this case,
the radial width of the bubble $\Delta_b$ is much smaller than the
radius of the bubble $R_b$, which means $\alpha\sim1$, and we can
express the bubble volume as $V=4\pi R_b^2 \Delta_b$. Now let's
estimate the radial width $\Delta_b$ at the time ($t_{sd}$) when
GRB-PWB interaction begins. Equating the ram pressure with the
internal pressure of the bubble, we have
\begin{equation}
\frac{1}{3}\bigg[\frac{(B^b)^2}{8\pi}+e_p \bigg]=\frac{L_w}{4\pi c R_s^2},
\end{equation}
where $e_p$ is the internal energy density of the particles.
The particle energy density can be regarded as the sum of the contributions from both the $e^+ e^-$ that
have suffered strongly cooling and those that have not suffered any cooling yet, i.e.
\begin{equation}
e_p=\bigg(\int_0^{t_{sd}-\tau_{syn}} \epsilon_e L_w\frac{\gamma_c^b}{\bar\gamma_e} \, dt+\int_{t_{sd}-\tau_{syn}}^{t_{sd}}
\epsilon_e L_w dt\bigg)/4\pi R_b^2 \Delta_b ,
\end{equation}
where $\gamma_c^b$ is the  Lorentz factor of  cooled $e^+ e^-$ in the bubble at time $t_{sd}$:
$\gamma_c^b=6\pi m_e c/(\sigma_T (B^b)^2 (t_{sd}-t))$.
Finally we get
\begin{equation}
\Delta_b=\bigg(\frac{6\pi m_e c}{\sigma_T (B^b)^2\epsilon_B
\gamma_w}{\rm ln}\frac{t_{sd}}{\tau_{syn}} +
\tau_{syn}\bigg)\frac{c}{3} \simeq 3\times 10^{16} {\rm cm}
\gamma_{w,5}^{-1}\epsilon_{e,0.5}^{-1}\epsilon_{B,0.5}^{-1}t_{sd,1}^3
{ \Delta{E}_{rot,53}^{-1}v_{b,0.1}^2 },
\end{equation}
consistent with our previous assumption $\Delta_b\ll R_b$.

The innermost part of the bubble is composed of newly injected
pairs from the pulsar that have suffered negligible cooling and
the minimum Lorentz factor is $\gamma_m^b$. For a nonmagnetic
ultrarelativistic shock, the downstream fluid just behind the
shock front moves with a velocity $v_d\simeq\frac{1}{3}c$ relative
to the shock front. The bulk of the shocked $e^+ e^-$ may move  at
a speed between $v_d$ and the bubble expanding speed $v_b$. As a
rough estimate, the width of these 'uncooled' $e^+ e^-$ region is
$\Delta_1\sim\bar{v}\tau_{syn}\sim 10^{16} {\rm cm}
\gamma_{w,5}^{-1}\epsilon_{e,0.5}^{-1}\epsilon_{B,0.5}^{-1}t_{sd,1}^3
{  \Delta{E}_{rot,53}^{-1}v_{b,0.1}^2}$, where we have taken $\bar
v$ being $0.2c$.

\section{GRB-PWB interaction and early GeV afterglows}
\subsection{Early GeV afterglows and GRB940217}

At the time $t_{sd}$ after the SN explosion, the pulsar collapses
to a black hole or converts to a strange star and the GRB goes
off, sending a fireball and a relativistic blast wave into the
PWB. Different from the usual cold proton-electron preshock medium
in the standard model of afterglows, the PWB is hot, composed of
randomly moving, relativistic $e^+ e^-$ pairs, and highly
magnetized. The initially freely-expanding fireball shell will be
decelerated by the swept-up hot $e^+ e^-$ pairs. We can calculate
the extent to which the shell is decelerated by PWB gas. The total
enthalpy, representing the effective inertia that decelerates the
shell, is $U\la4\pi R_b^2 \Delta_b (e_b+p_b)\simeq1.2\times10^{51}
{\rm erg}
\gamma_{w,5}^{-1}\epsilon_{e,0.5}^{-1}\epsilon_{B,0.5}^{-1}t_{sd,1}^2{
\Delta{E}_{rot,53}}$. If we temporarily neglect the radiative loss
of the GRB shell (i.e. the shell energy is conserved), its Lorentz
factor will be reduced from the initial value $\Gamma_0$ to
$\Gamma_f=(E_0^s/U)^{1/2}\simeq10(E^s_{0,53})^{1/2}
\gamma_{w,5}^{1/2}\epsilon_{e,0.5}^{1/2}\epsilon_{B,0.5}^{1/2}
t_{sd,1}^{-1}{ \Delta{E}_{rot,53}^{-1/2}}$ when the shell collides
with the outer supernova remnant, where $E_0^s=10^{53}{\rm
erg}E^s_{0,53}$ is the initial isotropic kinetic energy of the GRB
shell. The deceleration of the GRB shell starts when the energy of
the swept-up medium equals to $E_0^s/\Gamma_0^2$, lagging the GRB
prompt emission by a time $\Delta{t_{lag}}= R_b/2\Gamma_0^2 c=180
{\rm sec}\, t_{sd,1}{ v_{b,0.1}}\Gamma_{0,300}^{-2}$, where
$\Gamma_0=300\Gamma_{0,300}$ is the initial Lorentz factor. As the
forward shock expanding into the hot PWB,  the average Lorentz
factor per GRB-shocked $e^+ e^-$ particle is
$\bar\gamma^s_e=\epsilon_e \gamma_w \Gamma=1.5\times10^7
\epsilon_{e,0.5}\gamma_{w,5} \Gamma_{300}$, if the post
GRB-shocked medium has the same equipartition factors
$\epsilon_e$, $\epsilon_B$ as those in the pulsar wind shocks.
According to the shock jump conditions, the minimum Lorentz factor
of the GRB-shocked $e^+ e^-$ is
\begin{equation}
\gamma_m^s=\frac{p-2}{p-1}\epsilon_e \gamma_w \Gamma=2.5\times10^6 \epsilon_{e,0.5}\gamma_{w,5}\Gamma_{300}
\end{equation}
for $p=2.2$ and the magnetic field in the post GRB-shocked medium is
\begin{equation}
B^s=\Gamma B^b={ 51 {\rm Gauss}\,
\Gamma_{300}\epsilon_{B,0.5}^{1/2} t_{sd,1}^{-3/2}
\Delta{E}_{rot,53}^{1/2}v_{b,0.1}^{-1}.}
\end{equation}
The shock-accelerated electrons may have a maximum Lorentz factor
$\gamma_M$ determined by the balance between the shock accelerate
time (which could be the gyroperiod) and the radiative loss time.
Taking the shock-acceleration time  as the gyroperiod and assuming
synchrotron cooling , we get
\begin{equation}
\gamma_M^s\simeq (\frac{3\pi q_e}{\sigma_T B^s})^{1/2}= {
1.9\times10^7 \Gamma_{300}^{-1/2}\epsilon_{B,0.5}^{-1/4}
t_{sd,1}^{3/4} \Delta{E}_{rot,53}^{-1/4}v_{b,0.1}^{1/2}},
\end{equation}
where $q_e$ is the electron charge.   The observed characteristic
synchrotron emission { photon energy} for $e^+ e^-$ with the
Lorentz factor $\gamma_m^s$ is
\begin{equation}
h\nu_m^s=\Gamma (\gamma_m^s)^2 \frac{q_e B^s}{2\pi m_e c}={ 1.1
{\rm GeV} \,\Gamma_{300}^4 \gamma_{w,5}^2 \epsilon_{e,0.5}
\epsilon_{B,0.5}^{1/2}
t_{sd,1}^{-3/2}\Delta{E}_{rot,53}^{1/2}v_{b,0.1}^{-1} },
\end{equation}
and the maximum synchrotron emission { photon energy } is
\begin{equation}
h\nu_M^s=\frac{3 q_e^2}{m_e c \sigma_T}\Gamma=45 {\rm GeV}
\Gamma_{300} ,
\end{equation}
with the spectrum being $F_\nu\propto \nu^{-p/2}$ between them
{\footnote {Note that for a range of parameter values, the assumed
$\bar\gamma_e^s$ may become larger than $\gamma_M$, then one may
ask what will happens. To answer this,  one must know the detailed
microphysics of how the shock is formed. If there is a process
which thermalises particles faster than the gyroperiod, then this
process could lead to higher energy 'thermalised'  particles
rather than the 'nonthermal' maximum produced by the standard
Fermi acceleration picture (J. G. Kirk 2001, private
communication). In this case, the Fermi picture would not be
relevant. If there is no such a process, then the Lorentz factor
of the GRB-shocked $e^+ e^-$ pairs implied by the jump conditions
cannot be reached. The energy must be radiated away within the
shock structure itself (by synchrotron radiation) and the jump
conditions must be modified to account for the energy 'leak' (J.
G. Kirk 2001, private communication). Thus, strong high-energy
emissions are also expected.}}.

When the GRB blast wave swept-up matter cools, an initial GeV
burst occurs at the deceleration radius. { The deceleration of the
GRB shell starts when the energy of the swept-up medium equals to
$E_0^s/\Gamma_0^2$, and the corresponding  radial distance that
the shell  travelled in the PWB is $ x_{dec}=(E_s/\Gamma_0^2 4\pi
R_b^2(e_b+p_b))=3\times10^{13}{\rm
cm}\Gamma_{0,300}^{-2}t_{sd,1}$. Please note that, as the shell
slows down in the PWB, the radial distance $x$ that the shell
travelled is much shorter compared to its radius.  In this case,
 the angular spreading time scale }{\footnote {Because
of relativistic beaming, an observer sees up to solid angle of
$\Gamma^{-1}$ from the line of sight. Two photons emitted at the
same time and the same radius from the shell, one on the line of
sight and the other at an angle of $\Gamma^{-1}$ away, travel
different distance to the observer. The difference leads to a
delay in the arrival time by $\Delta t_{ang}= R_b/2\Gamma^2 c$ .}}
{ of the shell emission $\Delta t_{ang}= R_b/2\Gamma^2 c$ (Katz
1994a; Fenimore et al. 1996) dominates over the radial time scale
$\Delta t_{rad}=x/2\Gamma^2 c$, Thus an interesting situation
arise: though the time that the shell spends on moving through the
PWB is very short, the shock generated energy radiates over a much
longer observed time which is determined by the shell angular
spreading time. Therefore the shell angular spreading time is
relevant to the flux and light curve in the following calculation.
So, the effective  deceleration time should therefore be $t_{dec}=
R_b/2\Gamma_0^2 c=180 {\rm sec}
v_{b,0.1}t_{sd,1}\Gamma_{0,300}^{-2}$. }As the shocked PWB medium
is in the fast cooling regime, the power emitted is simply that
given to the shocked $e^+ e^-$, that is $\epsilon_e$ times the
power generated by the shock, i.e. $\int F_\nu d\nu= \epsilon_e
\frac{d E^s}{dt}\frac{1}{4\pi d_L^2} $ where $d_L$ is the
luminosity distance of the GRB source. Integrating the fast
cooling spectrum with $\nu_m^s\gg \nu_c^s$ over frequency (Sari et
al. 1998) , we have
\begin{equation}
\int F_\nu d\nu=\nu_m^s F_{\nu_m}^s \bigg[2+\frac{2}{p-2}(1-\bigg(\frac{\gamma_M^s}{\gamma_m^s}\bigg)^{2-p}\bigg]=
\epsilon_e \frac{d E^s}{dt}\frac{1}{4\pi d_L^2} ,
\end{equation}
where $E^s$ is the isotropic kinetic energy of the decelerated GRB
shell. { Strictly, to calculate the radiated power at the
deceleration radius, we need to calculate the total number of
swept-up relativistic electrons and their radiative power as in
 Dermer et al. 1999 . But as a rough estimate, we  assume that
the bulk of the shell kinetic energy is dissipated into the shock
energy at the deceleration radius (Rees \& M{\'e}sz{\'a}ros 1992;
Dermer et al. 1999) and get the fluence at $\nu_m^s\sim 1 {\rm
GeV}$ band (for $p=2.2$): }
%\begin{equation}
$t_{dec}\nu F_{\nu}(1{\rm GeV})=8\times10^{-6} {\rm erg
cm^{-2}}\epsilon_{e,0.5}E^s_{53}d_{L,28}^{-2}$ .
%\end{equation}
Clearly, this intensity of GeV emission is well above the detectability of EGRET detector.
This synchrotron GeV emission will fade with time and to a certain time it will be too weak to be detected,
leading to an extended  GeV afterglow emission.

Below we will study the fading behavior of GeV afterglow. {
According to Eq. (12), the observed afterglow flux at $\sim 1{\rm
GeV}$ is
\begin{equation}
\nu F_\nu(1{\rm GeV})=\nu_m^s F_{\nu_m}^s
(\frac{\nu}{\nu_m^s})^{-p/2+1} \propto (\nu_m^s)^{p/2-1}
\frac{dE^s}{dt}
\end{equation}
}

{  The duration of emissions from the shell scales with $\Gamma$
as $t\sim R_b/2\Gamma^2 c\propto \Gamma^{-2}$. }As the shocked gas
is partially radiative, the shock energy $E^s$ decreases as
$E^s\propto \Gamma^{\epsilon}$ in the limit that $\Gamma_0\gg
\Gamma \gg 1$ (B\"{o}ttcher \& Dermer 2000), where $\epsilon$ is
the fraction of the shock-generated thermal energy that is
radiated. For a fast-cooling shock, $\epsilon=\epsilon_e$. {
Considering  $\nu_m^s \propto \Gamma^4$, from Eq.(13) we find the
GeV spectral power flux decays with time as
\begin{equation}
\nu F_\nu(1{\rm GeV})\simeq 4.4\times10^{-8}{\rm erg cm^{-2}
s^{-1}}
(\frac{t}{t_{dec}})^{-(p-1)-\epsilon_e/2}\epsilon_{e,0.5}E^s_{53}d_{L,28}^{-2}t_{sd,1}^{-1}v_{b,0.1}^{-1}\Gamma_{0,300}^2.
\end{equation}
}
%we find
%\begin{equation}
%t\nu F_\nu(1{\rm GeV})=8\times10^{-6} {\rm erg cm^{-2}} (\frac{\Gamma}{\Gamma_0})^{2(p-2)+\epsilon_e} \epsilon_{e,0.5}
%E^s_{0,53}d_{L,28}^{-2}.
%\end{equation}
{ For comparison with EGRET observation and future {\it GLAST}
observation, we now calculate the energy fluence the detector can
collect during a certain time duration $t$ (see Fig.4 of Zhang \&
Meszaros 2001), i.e. the product of $\nu F_\nu$ Gev flux and the
observing time.}
\begin{equation}
t \nu F_\nu(1{\rm GeV})\simeq 8\times10^{-6}{\rm erg cm^{-2}
}\epsilon_{e,0.5}E^s_{53}d_{L,28}^{-2}
(\frac{t}{t_{dec}})^{-(p-2)-\epsilon_e/2}.
\end{equation}
The fluence threshold for EGRET is roughly $\sim2.1\times10^{-6}
{\rm erg cm^{-2}}$ for short integration time regime (Zhang \&
Meszaros 2001). { According to Eq. (15), the decreasing of the
fluence from the initial value at time $t_{dec}$ to the
sensitivity threshold of EGRET implies that  the observed GeV
emission  lasts}
\begin{equation}
t\sim  3500\,{\rm sec} \,
\epsilon_{e,0.5}^{2/(0.4+0.5\epsilon_{e,0.5})}(E^s_{0,53})^{2/(0.4+0.5\epsilon_{e,0.5})}
d_{L,28}^{-4/(0.4+0.5\epsilon_{e,0.5})}v_{b,0.1}t_{sd,1}\Gamma_{0,300}^{-2}.
\end{equation}
{ for $p=2.2$}. We think that this mechanism  provides a plausible
explanation for the famous long-duration ($\sim5400\,{\rm sec}$)
high-energy photons, with energies ranging from 36 Mev to 18 GeV,
from GRB940217 detected by EGRET (Hurley et al. 1994). The future
$GLAST$ detector is much more sensitive than EGRET and its
 fluence threshold   is roughly $\sim1.2\times10^{-9} t^{1/2} {\rm erg cm^{-2}}$
for a long integration time regime, so a more extended detected
GeV emission time is expected with $t_{dur}\sim 1.8\times10^{6}
{\rm sec}
\epsilon_{e,0.5}^{2/(1.4+0.5\epsilon_{e,0.5})}(E^s_{0,53})^{2/(1.4+0.5\epsilon_{e,0.5})}
d_{L,28}^{-4/(1.4+0.5\epsilon_{e,0.5})} \\
v_{b,0.1}t_{sd,1}\Gamma_{0,300}^{-2}$ for $p=2.2$ . The power-law
fading behavior of GeV emissions in our scenario distinguishes
itself from some other models, such as afterglow inverse-Compton
model ( Meszaros \& Rees 1994; Dermer 2000; Zhang \& Meszaros
2001) and hadron process models (e.g. Katz 1994b)

\subsection{ Synchrotron self-Compton  and nebula-induced external Compton components at GeV band}
We need to consider the synchrotron self-Compton (SSC) and
nebula-induced external Compton (EC) components at GeV band in
order to compare it with synchrotron component obtained in the
above subsection. The GRB-shocked $e^+ e^-$ gas will
inverse-Compton scatter synchrotron photons emitted by themselves
to higher energies. Because the shock is fast-cooling, the ratio
of the SSC to synchrotron luminosity is approximately given by
(Sari \& Esin 2001)

\begin{equation}
L_{SSC}=\left \{
       \begin{array}{ll}
         \frac{\epsilon_e}{\epsilon_B} L_{syn} \,\, \,\,\,\,\,\,\,\,\,\,\,\,\, {\rm if}~~~~ \epsilon_e\ll\epsilon_B \\
         \bigg(\frac{\epsilon_e}{\epsilon_B}\bigg)^{1/2} L_{syn} \,\,  {\rm if}~~~~\epsilon_e\gg\epsilon_B
        \end{array}
       \right.
\end{equation}
For fast-cooling shocks,  $L_{syn}\sim 4\pi d_L^2 \nu_m^s
F_{\nu_m}^s $ and $L_{SSC}\sim 4\pi d_L^2 \nu_m^{SSC}
F_{\nu_m}^{SSC}$, respectively. Note that
$\frac{\nu_m^s}{\Gamma}\gamma_m^s \gg m_e c^2$, the SSC energy
emission peaks at $\nu_{KN}=\Gamma \gamma_M^s m_e c^2 \\ \sim
3{\rm TeV} \Gamma_{300}^{1/2}\epsilon_{B,0.5}^{-1/4}
t_{sd,1}^{3/4}{ \Delta{E}_{rot,53}^{-1/4}v_{b,0.1}^{1/2}}$, above
which the Klein-Nishina effect suppresses Compton scattering.
Therefore, at $\sim 1{\rm GeV}$ band, the intensity of the SSC
fluence component is generally below that of the synchrotron one
so long as $\epsilon_e$ is not much larger  than $\epsilon_B$,
though at TeV band, it dominates over the latter.

Another important process that might contribute to the high-energy
component is the  Comptonization of external nebular soft photons
by relativistic $e^+ e^-$ in the expanding GRB shell. The
luminosity of the EC emission can be obtained analogously to the
SSC case, by using
%\begin{equation}
$\nu_m^{EC} F^{EC}_{\nu_m}\sim \tau_e \Gamma^2 \gamma_c^s
\gamma_m^s \nu_m^b F_{\nu_m}^b$
%\end{equation}
where $\tau_e$ is the optical depth of the shocked material. The
number density of $e^+ e^-$ in the bubble is $n_e^b=3\gamma_w
n_1={ 3L_w/4\pi\gamma_w R_s^2 m_e c^3=0.03
\gamma_{w,5}^{-1}t_{sd,1}^{-3}  \Delta{E_{rot}} v_{b,0.1}^{-2}}$,
so $\tau_e\sim 1/3\sigma_T n_e^b \Delta x\sim10^{-10}
\gamma_{w,5}^{-1}t_{sd,1}^{-3}{ \Delta{E_{rot}}
v_{b,0.1}^{-2}}\Delta{x_{16}}$, where $\Delta x$ is the radial
distance the shell travelled in the bubble. The luminosity of the
PWB $\nu_m^b F_{\nu_m}^b$ can be obtained from Eq. (9) in
Chevalier (2000): $\nu_m^b F_{\nu_m}^b\sim 10^{-14} {\rm erg
cm^{-2} s^{-1}}\epsilon_{e,0.5}^{1.1}\gamma_{w,5}^{0.2}
t_{sd,1}^{-1} d_{L,28}^{-2}$ for $p=2.2$. We finally get the EC
luminosity at time $t_{dec},$ $L^{EC}\sim 1.1\times10^{-12} {\rm
erg cm^{-2} s^{-1}}$, for typical parameters used . Noting that
$\gamma_m^s(\Gamma\nu_m^b)\gg m_e c^2$, the EC energy emission
peak also at $\nu_{KN}=\Gamma \gamma_M^s m_e c^2\sim 3{\rm TeV}
\Gamma_{300}^{1/2}\epsilon_{B,0.5}^{-1/4} t_{sd,1}^{3/4}{
\Delta{E}_{rot,53}^{-1/4}v_{b,0.1}^{1/2}}$. It is clear that EC
contribution to $\sim 1{\rm GeV}$ band is also unimportant
compared to the synchrotron component.

\subsection{ Attenuation of high-energy photons by soft photons in PWB}
As the PWB has plenty of soft photons, we must consider whether
the high-energy photons suffer strong attenuation due to
pair-production reaction while propagating through the PWB. The
optical depth of a high-energy photon with energy $E_\gamma$ due
to $\gamma-\gamma$ absorption writes
$\tau_{\gamma\gamma}=\int_{\nu_0}^{\nu_{max}}\sigma(\nu)n(\nu)\Delta_b
d\nu$ , where $h\nu_0\equiv 2(m_e c^2)^2/E_\gamma$ and $\nu_{max}$
is the  frequency of the maximum energy photon in the bubble,
 $n(\nu)\simeq L(\nu)/h\nu 4\pi R_b^2 c$ is the number density of
photons with energy $h\nu$ ($L(\nu)$ is the luminosity per unit
frequency of the PWB), and $\sigma(\nu)$ is the pair production
cross section of $E_\gamma$ with soft photon $h\nu$:
$\sigma(\nu)\sim\sigma_T (\frac{\nu}{\nu_0})^{-1}$ for
$\nu>\nu_0$.
%\begin{equation}
%\tau_{\gamma\gamma}\simeq \int_{\nu_0}^{\nu_{max}} \sigma_T \bigg(\frac{\nu}{\nu_0}\bigg)^{-1}\frac{L_\nu}{4\pi h \nu R_b^2 c}
%\Delta_b d\nu .
%\end{equation}
The formula of the PWB luminosity $L(\nu)$ has been derived by Chevalier (2000) (see his Eq. (9) ), given by
\begin{equation}
L(\nu)=\frac{1}{2} \bigg(\frac{p-2}{p-1}\bigg)^{p-1}\bigg[\frac{6q_e^2}{(2\pi m_e c)^2 c}\bigg]^{(p-2)/4}\epsilon_e^{p-1}
\epsilon_B^{(p-2)/4}\gamma_w^{p-2} R_s^{-(p-2)/2}L_w^{(p+2)/4}\nu^{-p/2} ,
\end{equation}
for $\nu>\nu_m^b>\nu_c^b$ and $L(\nu)\propto \nu^{-1/2}$ for
$\nu_m^b>\nu>\nu_c^b$. Setting $\tau_{\gamma\gamma}=1$, after a
simple algebraic computation, we get $h\nu_0\simeq 0.18 {\rm eV}
\epsilon_{e,0.5}^{0.18}\epsilon_{B,0.5}^{-0.86}\gamma_{w,5}^{-0.73}t_{sd,1}^{-0.14}{
\Delta{E}_{rot,53}^{0.045}v_{b,0.1}^{-0.09}}$ for $p=2.2$ .
Therefore the high-energy cutoffs due to pair production with soft
photons in PWB lie at
\begin{equation}
E_{cut}\simeq 5{\rm TeV}
\epsilon_{e,0.5}^{-0.18}\epsilon_{B,0.5}^{0.86}\gamma_{w,5}^{0.73}t_{sd,1}^{0.14}{
\Delta{E}_{rot,53}^{-0.045}v_{b,0.1}^{0.09}}
\end{equation}
for $p=2.2$. So we can safely conclude that the synchrotron GeV
photons can reach us without significant attenuation due to pair
production through interaction with soft photons in the PWB.

\section{Summary}

GRBs could arise within  PWBs in supranova-like scenarios (Vietri
\& Stella 1998; Wang et al. 2000) of GRBs, in which SN explosion
precedes GRB  by a time from several months to tens of years and
the pulsar emits a relativistic wind, making a wind bubble while
shocking against the ambient medium.
 After the spin-down time, the central pulsar may collapse to a black hole
or convert to a strange star, forming the GRB fireball. In this
paper, we studied the early afterglow high-energy emissions from
GRB blast wave expanding into such a pulsar wind bubble. We find
that, owing to that the PWB is composed of hot $e^+ e^-$ pairs
with high randomly-moving Lorentz factors, the majority of the
GRB-shocked PWB pairs have such high thermal Lorentz factors that
their characteristic
 synchrotron frequencies  lie at GeV bands.
Our calculation shows that the attenuation of  GeV photons by soft
photons in PWB due to pair-production reaction is negligible.
Strong power-law decaying GeV emissions during the early afterglow
phase is naturally expected in this scenario and it may provide a
plausible explanation for the long-time ($\sim 5400 {\rm s}$) GeV
photons from GRB940217.  The predicted power-law decaying behavior
of the flux is distinguished from the afterglow electron-IC
mechanism (Zhang \& M\'{e}sz\'{a}ros 2001) as the latter predicts
a delayed hump in the light curves.

\section*{Acknowledgments} { We are grateful to the referee for his valuable and careful comments}.
XYW  would like to thank  J. G. Kirk
for useful communication and Z. Li for valuable discussions. This
work was supported by the National Natural Science Foundation of
China under grants 19973003 and 19825109, and the National 973
project.

\end{document}